# Polaronic-Distortion-Driven Enhancement of Excitonic Auger Recombination in ZnO Nanoparticles

*Fuyong Hua,*[†,‡] *Zheng Zhang,*[†,‡] *Zhong Wang,*[†,‡] *Yang Liu,*[†] *Changchang Gong,*[†,‡] *Chunlong Hu,*[†,‡] *Yinhua Zhou,*[†] *Wenxi Liang\**[,†,‡]

[†]Wuhan National Laboratory for Optoelectronics, Huazhong University of Science and Technology, 1037 Luoyu Road, Wuhan, 430074, China

[‡]Advanced Biomedical Imaging Facility, Huazhong University of Science and Technology, Wuhan, 430074, China

**ABSTRACT:** The surface effect and quantum confinement render nanomaterials the optoelectronic properties more susceptible to nonradiative processes than their bulk counterparts. These nonradiative processes usually contain a series of interwoven and competing sub-processes, which are challenging to disentangle. Here, we investigate the structural origin of Auger recombination in ZnO nanoparticles using transient absorption spectroscopy and ultrafast electron diffraction. The photogenerated hot holes are captured by oxygen vacancies through an Auger mechanism, inducing significant local structural distortions around the oxygen vacancy and its neighboring zinc tetrahedron on a sub-picosecond timescale. The recombination of trapped holes accelerates the lattice thermalization and stabilizes the formed small hole polarons. Subsequently, the recombination of localized polarons forms a confined exciton–polaron complex that may account for the long-lived (>7 ns) visible luminescence observed in ZnO nanoparticles. Our findings are potentially applicable to other transition metal oxide nanomaterials, bringing insights for the optimization of their functional properties.

Zinc oxide nanoparticles (ZnO-NPs) are widely adopted in many applications, for example, as the electron transport layer in perovskite solar cells for their many advantages,[1] including the favorable optoelectronic properties and the potential of large-scale printing production.[2,3] Undoped ZnO nanomaterials possess an intrinsic n-type conductivity, which is attributed to the abundant defect sites, particularly, to the oxygen vacancies near surface.[4,5] Although a wide range of defect types have been identified,[6] the underlying mechanisms of defect-mediated carrier relaxation and transport are not yet fully elucidated. Using oxygen vacancy, one of the major species, as an example, the temperature-dependent measurements demonstrated that the luminescence in visible range was associated with the recombination of electrons at the shallow defect levels of doubly ionized oxygen vacancies, $V_O^{++}$, which were generated through a two-step hole capturing process of $V_O \rightarrow V_O^+ \rightarrow V_O^{++}$;[7,8] The First-principles calculations revealed the deep donor levels formed by these vacancies,[9] and the local lattice distortion up to 23% of the Zn-O bond length.[10] However, so far no experimental observation on structural evolution testifies these scenarios.

To address the role of oxygen vacancies, pump-probe based techniques have been employed in the investigations of defect-associated transitions in ZnO-NPs. For example, the transient absorption (TA) spectroscopy revealed the hopping carrier diffusion within the energetically continuous defect states,[11] and identified the polaron excited state that injected an additional electron into the conduction band (CB) with prolonged lifetime.[12] The time-resolved X-ray photoelectron spectroscopy confirmed that the hole capture by oxygen vacancies induced a local structural distortion corresponding to 15% of the Zn-O bond length within 80 ps,[13] consistent with the computational simulations.[10] These observations obtained only, however, the carrier-derived information through differential spectroscopy, indirectly inferring the impact of structural responses on the process of defect capturing. Investigations into atomic motions are essential to elucidate the complex carrier–lattice interactions related to the self-trapping and polaron formation in ZnO-NPs. Compared to TA, which probes the changes of the joint density of carrier states[14], ultrafast electron diffraction (UED) captures the lattice evolutions in reciprocal space,[15] e.g., phonon–phonon scattering,[16] anisotropic atomic motions,[17] structural distortions,[18] and so on.[19–21] The sensitivity of direct structure information makes UED a suitable approach to assess the carrier capture and lattice deformation in ZnO-NPs.

In this work, we investigate the formation of exciton–polarons in ZnO-NPs by combing the complementary probes of TA and UED, to uncover the structural origin underlying the polaronically enhanced excitonic Auger recombination. Upon photoexcitation, the surface oxygen vacancies capture the photogenerated holes within 200 fs, forming the small polarons that subsequently trap the additional electrons in CB to create the long-lived excitonic polarons. This process is accompanied by the local lattice distortion and the compression of Zn-O bonds, which are associated with the oxygen vacancy, enhancing the electron–phonon scattering and promoting the lattice thermalization through an Auger-mediated mechanism. At the end, such a polaron-associated relaxation pathway strongly impacts the overall energy relaxation in ZnO-NPs.

The investigated wurtzite-phase ZnO-NPs were synthesized via a methanolic sol-gel/precipitation method, and characterized by high-resolution transmission electron microscopy (HRTEM) and optical spectroscopies (see Experimental Section and Figure S1 in Supporting Information). The HRTEM images reveal the average particle diameter of 6–7 nm and the hexagonal lattice structure. The steady-state absorption spectrum exhibits a clear absorption edge, from which the bandgap is extracted as ~3.36 eV. The photoluminescence (PL) spectrum features two discrete emission peaks centered at 3.1 eV ($PL_{UV}$) and 2.3 eV ($PL_{VIS}$), constituting 5.1% and 94.6% of the total intensity, respectively. The former largely arises from the recombination of band-edge excitons[22] or localized excitons within the defect states close to band edge,[11] while the latter is primarily attributed to the luminescence associated with self-trapped excitons and defects.[23]

We first examine the carrier relaxation upon above-bandgap excitation of 320 nm using TA spectroscopy in transmission geometry. As shown in Figure 1a, the TA spectrum exhibits two photo-induced absorption signals at 2.4–3.0 eV (PIA1) and 3.4 eV (PIA2), and a photo-bleaching (PB) signal at 3.0–3.4 eV. The broadband PIA1 and PB dominate the early few picoseconds, then diminish significantly in ~50 ps, leaving only PIA2 that lasts more than 7 ns, as depicted in Figure 1b.

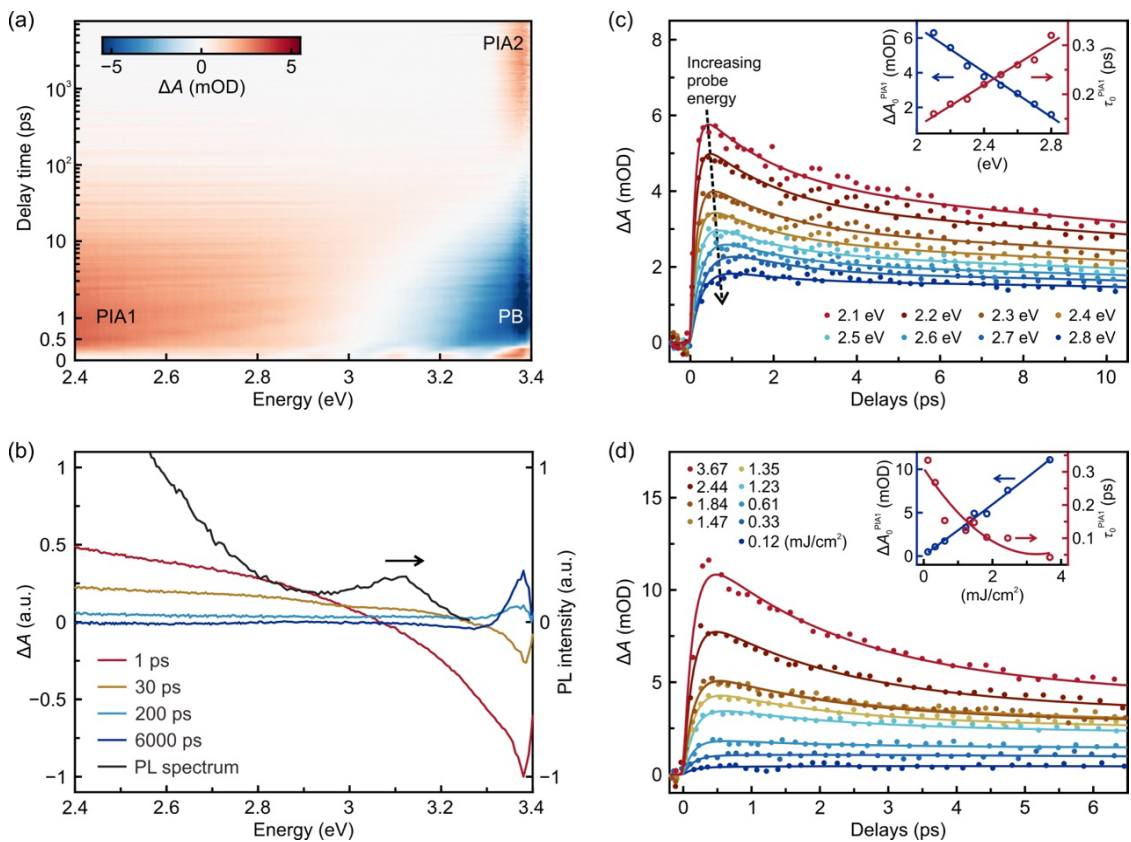

**Figure 1.** Hole capturing revealed by TA measurements. (a) Pseudocolor contour plot of the TA result upon excitation with fluence of 1.23 mJ/cm². (b) Spectra extracted from (a) at selected delay times, shown with the PL spectrum for comparison. The blue shift between the PB peak and $PL_{UV}$ may arise from the large exciton binding energy.[24,25] (c) PIA1 kinetics probed at various energies, extracted from (a). Solid lines, multiexponential fits. Inset, dependences of $\tau_0^{PIA1}$ and $\Delta A_0^{PIA1}$ on probe energy; Solid lines, linear fits. (d) PIA1 kinetics at 2.4 eV measured with various fluences. Solid lines, multiexponential fits. Inset, dependences of $\tau_0^{PIA1}$ and $\Delta A_0^{PIA1}$ on excitation fluence; Solid lines, the guides to the eye. See Table S1 and S2 in Supporting Information for the results of multiexponential fit.

The rise time of PIA1 ($\tau_0^{PIA1}$) increases with increasing probe energy while the maximum amplitude ($\Delta A_0^{PIA1}$) linearly decreases, as depicted in Figure 1c. We fit the PIA1 kinetics probed at various energies, yielding rise times of 160–320 fs, which align with the reported self-trapping times.[23,26] Such a probe-energy-dependent rise suggests the carrier relaxation from the shallow defect levels to the deep ones after defect trapping,[26,27] see Supplementary Discussion 1 in Supporting Information. Considering the high mobility of electrons in ZnO-NPs,[28,29] holes are expected to be the major species of trapped carriers.[30] This picture is corroborated by the fluence-dependent measurements, which show that $\Delta A_0^{PIA1}$, probed at 2.4 eV, scales linearly with increasing fluence without noticeable saturation, while $\tau_0^{PIA1}$ decreases dramatically from 330 fs to 42 fs, as depicted in Figure 1d. This expedited rise is interpreted by the increased population of electron–hole pairs, which accelerate the lattice thermalization via enhanced Auger trapping[31,32] due to the quantum confinement effects upon strong excitation.[33] Moreover, the high exciton density in the initial stage after photoexcitation also induces an excitonic Auger recombination, resulting in the increased rate of hole capturing.[26,34]

The band-like PB signal extending in 3.0–3.4 eV, which shares the same origin with the Urbach tail in steady-state absorption spectrum (see Figure S1c in Supporting Information), is jointly contributed by the continuous defect states near band edge and the states above CB. The rise of PB band ($\tau_0^{PB}$) occurs in 100–200 fs (see Table S3 in Supporting Information) is similar to $\tau_0^{PIA1}$, indicating the simultaneous excitation and capturing of both hot holes and electrons. In contrast, the decay of PB band ($\tau_1^{PB}$) slows with increasing probe energy, as depicted in Figure 2a. This trend further corroborates the picture of defect trapping, in which the defect states closer to band edge initially capture more electrons, resulting in the longer time spans of decay via recombination.[11,35] As the probe energy approaches the band edge (~3.4 eV), the bleach signal overshoots sharply after the delay time of ~100 ps, giving rise to PIA2.

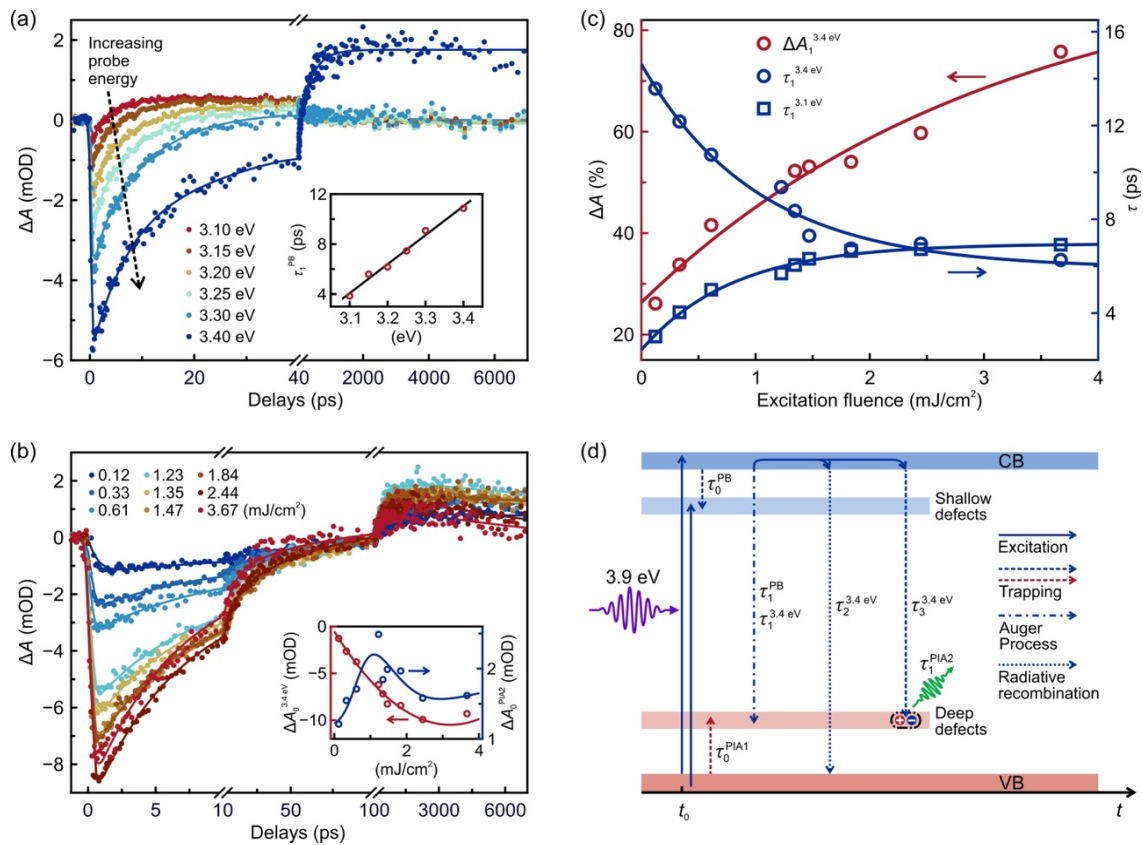

**Figure 2.** Bleach signal evolution and the carrier relaxation processes. (a) Dependence of PB kinetics on probe energy. Solid lines, multiexponential fits. Inset, linear dependence of $\tau_1^{PB}$ on probe energy.

(b) Dependence of band-edge PB kinetics on excitation fluence. Solid lines, multiexponential fits. Inset, dependences of $\Delta A_0^{3.4\ eV}$ and $\Delta A_0^{PIA2}$ on fluence; Solid lines, the guides to the eye. See Table S3 and S4 in Supporting Information for the results of multiexponential fit. (c) Fitted characteristic parameters of PB kinetics at 3.1 and 3.4 eV as a function of fluence. Solid lines, the guides to the eye. (d) Schematic illustration of carrier relaxation in ZnO-NPs, see text.

The band-edge PB peak induced by the bleach of free excitons,[36] demonstrates a saturation of bleach amplitude ($\Delta A_0^{3.4\ eV}$) upon excitation of 2.44 mJ/cm$^2$, as shown in Figure 2b. Furthermore, the decay of band-edge PB shows a complex evolution. As depicted in Figure 2c, the first decay amplitude ($\Delta A_1^{3.4\ eV}$) increases with increasing fluence, agreeing with the enhanced Auger recombination in nanoparticles[37] and the well-studied non-radiative recombination in ZnO nanomaterials.[36,38,39] Upon excitation of low fluence, the first decay time, $\tau_1^{3.4\ eV}$, is significantly longer than that of PB band at 3.1 eV, $\tau_1^{3.1\ eV}$ (see Figure S2 and Table S5 in Supporting Information for details). $\tau_1^{3.4\ eV}$ is prolonged due probably to the long-lived CB electrons injected from the excited polaronic states with limited recombination channels,[12] and the non-radiative recombination of the free excitons with the holes trapped in deep defects.[36,40] As we increase the excitation fluence, $\tau_1^{3.4\ eV}$ and $\tau_1^{3.1\ eV}$ change in opposite ways until the fluence reaches the shared saturation threshold of 2.44 mJ/cm$^2$, see Figure 2c. The decreasing $\tau_1^{3.4\ eV}$ also agrees with the picture of enhanced Auger recombination, as the increased carrier density accelerates many-body interactions in confined nanoparticles.[33] This trend together with the saturation of $\Delta A_0^{3.4\ eV}$ suggest a scenario of exciton absorption with photoinduced transparency, in which the exciton relaxation near band edge probably transitions from the exciton–phonon scattering to the exciton–exciton scattering, or to the electron–hole plasma generation.[39] During this process, the constraint of defect-assisted Auger recombination[41] results in the increasing $\tau_1^{3.1\ eV}$.[36] Upon excitation with saturation fluence, the recombination rates are governed by the Auger processes, leading to the convergence of $\tau_1^{3.1\,eV}$ and $\tau_1^{3.4\,eV}$, regardless of the carrier origin of exciton or defect state. The following decay (characterized by $\tau_2^{3.4\,eV}$) in several tens of picoseconds is assigned to the radiative recombination, and the much slower decay (characterized by $\tau_3^{3.4\,eV}$) in several hundreds of picoseconds is corresponding to the relaxation into deep defect states, see Supplementary Discussion 2 in Supporting Information.

In contrast, PIA2 demonstrates a non-monotonic trend of maximum amplitude ($\Delta A_0^{PIA2}$) change with increasing fluence, with a saturation at 1.23 mJ/cm$^2$ (see the inset of Figure 2b), reflecting the recombination of defect-bound excitons which possess a long-lived lifetime ($\tau_1^{PIA2}$) more than 7 ns,[26,42] see Supplementary Discussion 3 in Supporting Information.

At this point we are able to summarize the carrier relaxation dominated by defect trapping and Auger processes, as illustrated in Figure 2d. Upon above-bandgap excitation, hot holes are captured by the deep defect states, which are associated with oxygen vacancies, while hot electrons fill the CB edge and shallow defect states, in the similar time spans of $\tau_0^{PIA1}$ and $\tau_0^{PB}$, respectively. Then the Auger recombination processes, with defects involved, occur during $\tau_1^{PB}$ and $\tau_1^{3.4\,eV}$, followed by the radiative recombination of re-excited extra electrons within $\tau_2^{3.4\,eV}$. For the remaining electrons in CB, a slow capturing by the deep defects occurs in $\tau_3^{3.4\,eV}$, resulting in the formation of exciton polarons through binding with the previously localized holes. Subsequently, the recombination of exciton polarons yields the green luminescence with a significantly extended lifetime of $\tau_1^{PIA2}$. Both the polaron formation and the subsequent evolutions are associated to lattice distortions, as discussed next.

The lattice distortion is revealed by UED measurements (see Experimental Section in Supporting Information). Across the probed scattering vector $Q$, the radially averaged intensity of all indexed diffraction peaks decreases upon excitation of 343 nm, indicating the lattice thermalization and the increase of root-mean-square atomic displacement,[16] as shown in Figure 3a. Particularly, the (002) and (013) peaks exhibit derivative-like evolutions within 0–20 ps (see Figure S3 in Supporting Information), suggesting a local carrier–lattice interaction,[43] which is a precursor phenomenon often linked to polaronic effects.[44]

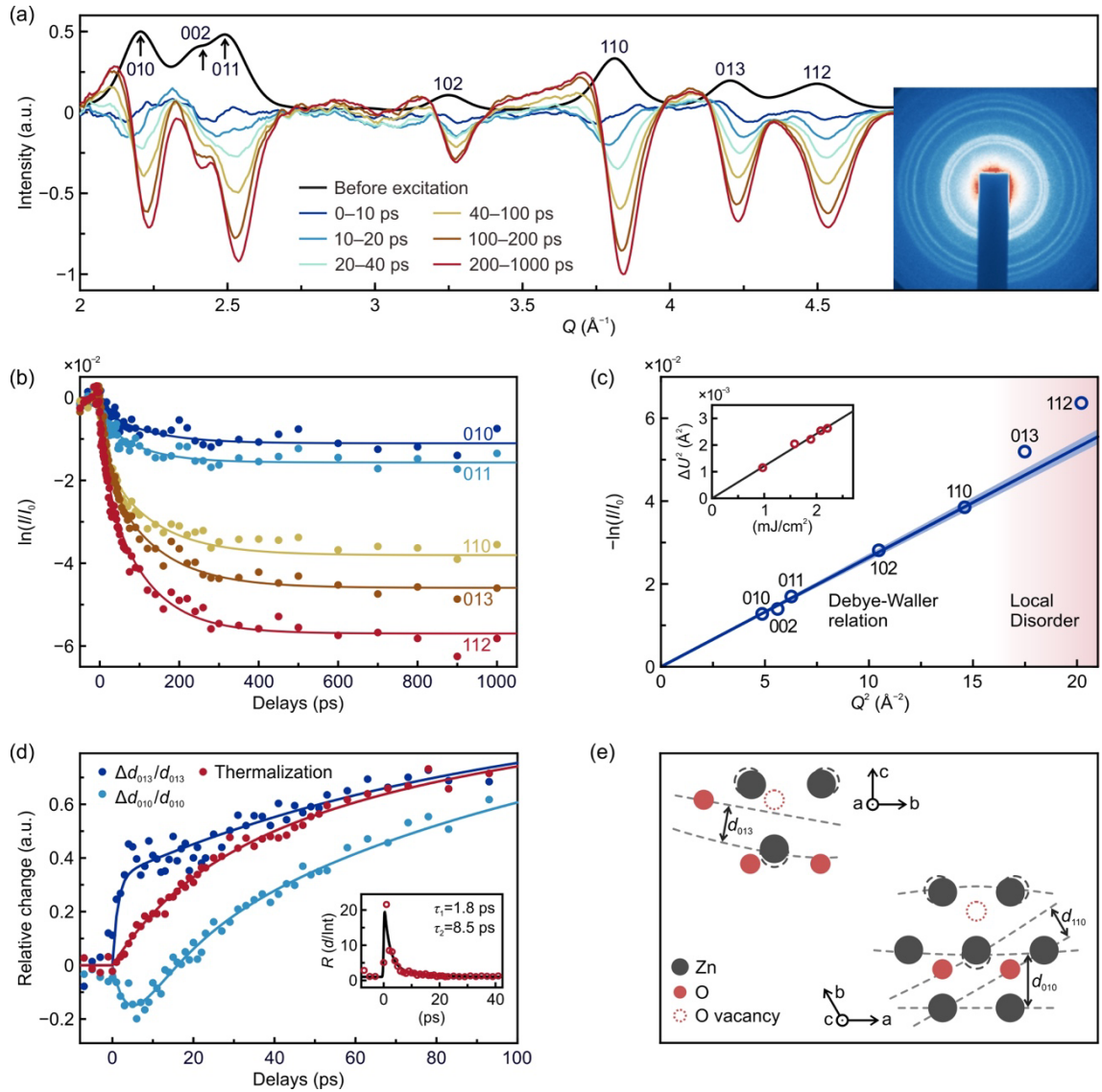

**Figure 3.** Lattice dynamics of global thermalization and local structural distortion. (a) Differential diffraction intensity at selected delay times, excited with 2.21 mJ/cm$^2$. Black line, background-subtracted intensity profile. Inset, recorded pattern. (b) Temporal evolutions of relative intensity change of the selected diffraction peaks with high signal-to-noise ratio. $I_0$, intensity measured before excitation.

Solid lines, biexponential fits (see Table S6 in Supporting Information for fit results). (c) Dependence of the intensity changes at delay time of 1000 ps on squared scattering vector. Inset, dependence of root-mean-square atomic displacement on fluence, derived from the (010), (002), (011), (102), and (110) peaks. (d) Comparison of the interplanar spacing changes of (013) and (010) to the inversed averaged intensity change (thermalization, see text). All traces are normalized. Inset, temporal evolution of distortion ratio. (e) Schematic illustration of the interplanar spacing changes contributed from zinc atom motions. Filled balls, atomic sites at equilibrium.

The intensity changes of diffraction peak show two-stage decays with characteristic times of ~17 ps and ~100 ps, as depicted in Fig. 3b. The rapid process is comparable to those observed in other quantum dots,[18,31,45] reflecting the energy transfer from hot excitons to lattice via phonon emission and Auger recombination,[31,46] i.e., carrier–phonon coupling. The slow process involves the thermal equilibration within phonon system and the heat conducting induced by the carbon film substrate.[47] We quantified the atomic displacements at equilibrium under the Debye-Waller description, $-\ln(I/I_0) = \Delta\langle u^2\rangle_{\Delta T} \times Q^2$,[16] as depicted in Fig. 3c. The high-order peaks of (013) and (112) deviate noticeably from the linear dependence on squared scattering vector, $Q^2$, suggesting local structural distortions besides the global thermal fluctuations.[13,18,40] Such deviations are indicative of the expected behavior of small polarons driven by the hole trapping through surface defects, see Supplementary Discussion 4 in Supporting Information.

Besides the intensity changes, the interplanar spacing changes also encode the information of local distortion. We employed the averaged intensity change of (013) and (112) peaks, which show identical evolutions (see Figure S5 in Supporting Information), as the assessment of global lattice thermalization.[48] As the thermalization process shows a biexponential rise with characteristic times of 16.7 ps and 92 ps, the interplanar spacing of (010) and (013) show distinct evolutions, in which the (013) plane exhibits a rapid expansion in 1.3 ps while the (010) plane exhibits an initial compression in the same duration, see Fig. 3d. Both planes then undergo a much slower expansion with characteristic times of 95 and 106 ps, respectively, aligning with the slow thermalization rate. We calculated the ratio of the interplanar spacing change of (013) over the global thermalization, denoted as $R(d/\text{int})$, to quantify the distortion rate, as depicted in the inset of Fig. 3d. Fitting the trace of $R(d/\text{int})$ yields a rise time of 1.8 ps and a decay time of 8.5 ps, corresponding to the establishment of local distortion and the recovery of distortion towards the global thermalization, respectively (see Supplementary Discussion 5 and 6 in Supporting Information for more discussions). The 1.8 ps rise time closely matches the duration of hole diffusion and trapping associated to the singly charged oxygen vacancies in ZnO-NPs.[40] In contrast, the identified hot hole capturing process with $\tau_0^{\text{PIA1}} \sim 200$ fs in our TA measurement, appears apparently faster due to the sensitivity of TA spectroscopy to the carrier cooling and exciton formation, both of which occur on a sub-picosecond timescale. The subsequent stabilization of excitons by local lattice distortion is more prominent when it is observed from the perspective of structural dynamics.[40] With further inspecting the time spans of anomalous interplanar spacing change (see Supplementary Discussion 7 in Supporting Information), we are able to conclude that the preceding time of distortion reflects the formation and stabilization of excited-state hole-polarons.[12,40,49]

As schematically illustrated in Fig. 3e, the reflection of closely spaced (013) planes is particularly sensitive to the displacement of zinc atoms along the $c$-axis, while the (110) plane spacing primarily captures the basal-plane distortions. When holes are captured, the separations between the oxygen vacancy and its nearest four zinc atoms increase resulting in the observed rapid expansion of (013) plane. Due to the large projections of normal vector lying in the $a$-$b$ plane, the correlated (010) and (110) planes experience compression when the four zinc atoms repel each other during hole capturing,[10] as manifested by the spacing decrease of (010) shown in Fig. 3d. The reported decrease from 0.87 to 0.79 for the ratio of Young's moduli in the direction of *a*-axis over that in the *c*-axis after the hole capturing by oxygen vacancies,[50] probably facilitates a broader compressive response in the *a-b* plane, hence making the compressive deformation along the *c*-axis less possible.

Further details of atomic motions involved in the formation and relaxation of hole polarons are inspected by the analysis of pair-distribution function (PDF). The calculated differential PDF, $\Delta G(r,t)$, accesses four distinct atomic pairs in a distance range of 10 Å, as depicted in Fig. 4a. The peak at 1.9 Å ($G_{1.9\text{Å}}$) corresponds to the first (nearest-neighboring) Zn-O pairs, representing the interactions between the oxygen atom and its surrounding zinc atoms. The 3.3 Å peak ($G_{3.3\text{ Å}}$) corresponds to the first Zn-Zn pairs, representing the separation of zinc atoms impacted directly by the oxygen vacancy. These two peaks reflect the local atomic environment near oxygen vacancies. In contrast, the peaks at 6.0 Å ($G_{6.0\text{ Å}}$) and 8.7 Å ($G_{8.7\text{ Å}}$) correspond to the fifth and the seventh neighboring Zn-Zn pairs with large distance, respectively, reflecting the lattice-scale atomic motions, e.g., the global lattice thermalization. The profiles of $\Delta G(r,t)$ at selected delay times (see Fig. 4a) and the intensity changes of four peaks (see Fig. 4b) reveal that, the peaks with short distance undergo more pronounced intensity reduction than those with large distance, due probably to the local distortions perturbing the uniform lattice heating. The local distortions modify the interatomic separations of low-order atomic pairs, leading to the deviations from Debye-Waller effect during thermalization,[18] consistent with the behavior of high-indexed diffraction peaks shown in Fig. 3c.

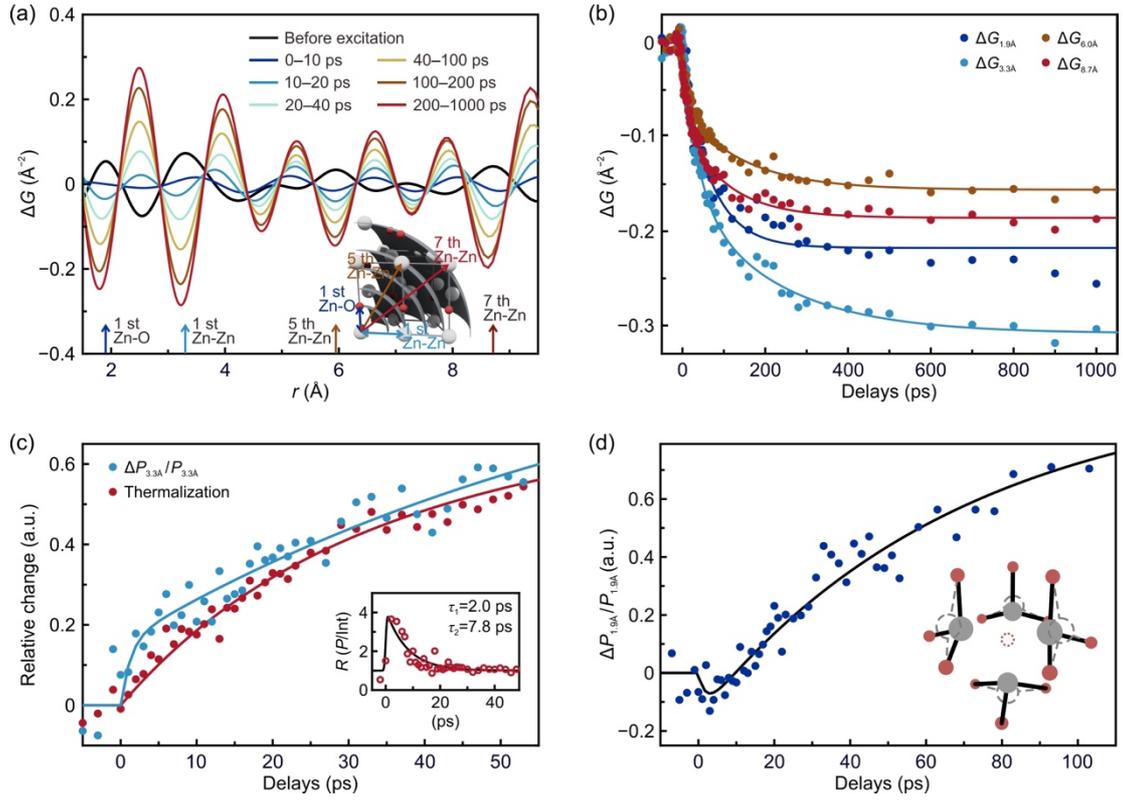

**Figure 4.** Analysis of pair distribution functions. (a) Calculated differential PDFs at selected delay times upon excitation of 2.21 mJ/cm$^2$. Inset, schematic of atomic shell structure of wurtzite-phase ZnO marked with the resolved atom pairs. (b) Kinetics of $\Delta G$ of four atom pairs in (a). Solid lines, biexponential fits. (c) Comparison of the position change to the inversed intensity change (thermalization, see text) for the first zinc-zinc pairs, $G_{3.3\text{Å}}$. Solid lines, biexponential fits. Inset, temporal evolution of the ratio of $P_{3.3\text{Å}}$ over lattice thermalization. (d) Initial compression of the first zinc-oxygen pair followed by a thermal expansion. Inset, schematic illustration of the outward displacement of zinc atoms due to the charge repulsion induced by the captured hole at oxygen vacancy site.

The temporal change of PDF peak position corroborates the observed evolution of interplanar spacing. Similar to the kinetics of (013) interplanar spacing (see Fig. 3d), the relative distance change of the nearest Zn-Zn pairs ($G_{3.3\text{Å}}$ peak position, $P_{3.3\text{Å}}$) shows a rapid expansion in 1.6 ps, preceding the decrease of peak intensity that also reflects the lattice thermalization, as depicted in Fig. 4c. The expansion rate significantly surpasses the rate of lattice thermalization, suggesting the formation of oxygen-vacancy polarons, in which the adjacent zinc atoms move outward when an oxygen vacancy captures a hole, resulting in the increase of the nearest Zn-Zn pair separation.[13] We again calculated the ratio of $P_{3.3Å}$ over thermalization, which shows similar evolution to that of (013) plane, as depicted in the inset of Fig. 4c, again corroborating the scenario of polaron formation and stabilization.

We also examined the evolution of first Zn-O pairs ($G_{1.9Å}$), finding an initial reduction of distance similar to the kinetics of (010) interplanar spacing, as depicted in Fig. 4d. This reduction can be interpreted as the zinc atoms squeeze towards the adjacent oxygen atoms during they displace away from the hole-captured vacancy. Such a coupling effect enables the spin-forbidden singlet-triplet transitions, as discussed in Supplementary Discussion 2 in Supporting Information, facilitating the radiative transitions of polaronic excitons and resulting in the green luminescence that characterizes ZnO-NPs.[51]

In summary, we resolved the pathway and time spans of the formation of exciton-polarons in ZnO-NPs, through the observations in both degrees of freedom of charge and lattice. Our observations highlight the occurrence of local distortion around oxygen vacancy and its neighboring zinc tetrahedron in ~1 ps, followed by the lattice stabilization through the cooperative Zn-O bond rearrangement within several picoseconds. Thus, the structure relaxes into a lower energy state, forming the excited-state polarons, which stabilize into small hole polarons. The stabilized polarons[52] can recombine with the electrons in CB via Auger capturing or defect-assisted processes, resulting in the confined exciton states. Furthermore, these long-lived polarons continuously perturb the lattice and serve as a key-mechanism underlying the persistent green luminescence.

Strongly polar metal oxide materials like ZnO, are particularly susceptible to the intrinsic defect formation that generates a substantial population of small polarons,[53] laying significant impacts on the material properties in many ways, e.g., enhancing photogenerated carrier mobility,[54] modifying exciton binding energies, and inducing polaronic absorption bands and band bending.[55] On the other hand, the strong electron–phonon interactions associated with small polarons can facilitate carrier localization and mediate the non-radiative recombination, by providing additional relaxation pathways, thereby suppressing radiative emission[56] and quenching visible emission,[57] which are not favored by light emitting applications. Our findings bring better understanding of small polaron formation and relaxation dynamics in ZnO-NPs, and possibly in nanomaterials with alike structure, providing essential information for designing strategies to effectively optimize the material's performance.

## ASSOCIATED CONTENT

**Supporting Information**

Experimental details of sample preparation, characterization, transient absorption spectroscopy, and ultrafast electron diffraction measurements; Discussions on the origins of visible TA signals, band-edge bleaching dynamics, defect-related absorption, and lattice distortions linked to Auger-induced thermalization and small-polaron formation; Additional figures S1–S7, tables S1–S7. (PDF)

## AUTHOR INFORMATION

**Corresponding Author**

* Email: wxliang@hust.edu.cn**Author Contributions**

W.L. conceived of and supervised the project. F.H. performed the measurements with the supports from Z.Z., Z.W., C. G. and C. H. Y.L synthesized the ZnO nanoparticles under the supervision from Y.Z. F.H. and W.L. analyzed the data with discussions with all authors. F.H. and W.L. wrote the paper with the contributions from all authors.

**Acknowledgement**

We thank the Analytical and Testing Center in Huazhong University of Science and Technology for the supports. Y.L. and Y.Z. thank the support from the National Natural Science Foundation of China (52452307).

*Supporting Information for*

# Polaronic-Distortion-Driven Enhancement of Excitonic Auger Recombination in ZnO Nanoparticles

Fuyong Hua,[†,‡] Zheng Zhang,[†,‡] Zhong Wang,[†,‡] Yang Liu,[†] Changchang Gong,[†,‡] Chunlong Hu,[†,‡] Yinhua Zhou,[†] Wenxi Liang*,[†,‡]

[†]Wuhan National Laboratory for Optoelectronics, Huazhong University of Science and Technology, 1037 Luoyu Road, Wuhan, 430074, China

[‡]Advanced Biomedical Imaging Facility, Huazhong University of Science and Technology, Wuhan, 430074, China

*Corresponding Author. Email: wxliang@hust.edu.cn

**Table of Contents:**

I. **Experimental section**

**Sample preparation**

The ZnO nanoparticles were synthesized following a methanolic sol-gel/precipitation method adapted from Spanhel and Anderson.[1] In a typical procedure, zinc acetate dihydrate ($Zn(CH_3COO)_2 \cdot 2H_2O$) was dissolved in methanol to form solution A (0.03–0.1 M). Separately, potassium hydroxide (KOH) was dissolved in methanol to form a basic solution B (KOH:Zn = 2–3 mol ratio). Solution A was added dropwise into solution B within 10 min under vigorous stirring at room temperature (water-bath at 25–30 °C). The mixture was stirred for 2–3 hours until the reaction completed, and then allowed to settle. The resulting white precipitate was collected by centrifugation, then washed three times with methanol, and finally redispersed in a 7:3 (v/v) mixture of n-butanol and chloroform. The obtained ZnO nanoparticles typically exhibit diameters below 10 nm, with the size distribution dependent on the Zn:KOH ratio, adding rate, and reaction temperature.

The obtained solution of ZnO nanoparticle was subsequently diluted tenfold with methanol to prepare samples for the experiments of carrier and structure dynamics observation. In transient absorption measurements, the diluted solution was transferred into a 1 mm-thick optical cuvette for the spectroscopic examination. In ultrafast electron diffraction measurements, a dip-coating method was employed, in which a carbon-coated copper grid was vertically withdrawn from the diluted solution at a speed of 0.05 mm/s. This operation allowed the formation of a monolayer of uniformly dispersed ZnO nanoparticles on the surface of carbon surface.

**Sample characterization**

Figure S1a and S1b show the morphology of the as-prepared ZnO nanoparticles deposited on carbon-coated copper grid, which is imaged by the high-resolution transmission electron microscopy (HRTEM, JEM-ARM 200F, JEOL). An average radius of 6–7 nm is observed. The nanoparticles are uniformly distributed across the substrate surface, showing no significant aggregation or overlapping. The fast Fourier transform (FFT) of the HRTEM image confirms the hexagonal wurtzite crystal structure, indicating a high degree of crystallinity.

Figure S1c shows the steady-state absorption spectrum (Lambda 35, PerkinElmer). The Tauc plot yields an optical bandgap of 3.36 eV, with the Urbach tail clearly visible in the sub-bandgap range. Figure S1d shows the photoluminescence spectrum (QuantaMaster 8000, HORIBA) excited at 340 nm. The Gaussian fit yields two peaks centered at 2.3 eV and 3.1 eV, denoted as $PL_{VIS}$ and $PL_{UV}$, respectively.

**Transient absorption spectroscopy**

The transient absorption (TA) measurements in transmission geometry were performed under ambient conditions using a Helios spectrometer (Ultrafast Systems), driven by a Ti:sapphire regenerative amplifier (Legend, Coherent Inc.) operating at a repetition rate of 5 kHz and delivering ~35 fs pulses centered at 800 nm. The fundamental output was directed into a TOPAS optical parametric amplifier (Light Conversion Inc.) to generate 320 nm pump pulses. The pump beam was mechanically chopped to 2.5 kHz before exciting the sample, while a white-light continuum spanning the UV–visible range was generated by focusing a small portion of the 800 nm beam into a $CaF_2$ crystal, serving as the probe. The diameters of the pump and probe spots on sample were set to ~200 μm and ~60 μm, respectively. The instrument response function of the setup was measured to be ~100 fs.

**Ultrafast electron diffraction**

The ultrafast electron diffraction (UED) measurements were performed under ultrahigh vacuum conditions using a home-built ultrafast electron diffractometer, which temporal resolution was estimated to be less than 1 ps. The samples were excited by 343 nm pulses, which were the third harmonic of the 1028 nm output from femtosecond laser (CARBIDE, Light Conversion), with a repetition rate of 14.3 kHz. The probe electron pulses were accelerated to 30 keV, impinging in a transmission geometry. Diffraction patterns were intensified by a microchannel plate detector (VID240, Photek) then recorded by an sCMOS camera (ORCA flash 4.0, Hamamatsu), with $1.43 \times 10^5$ pulses accumulated for each frame. The diameters of the pump and probe spots on sample were set to ~600 μm and ~100 μm, respectively. Time-resolved diffraction signals were radially integrated to yield the scattering-vector-dependent intensity profiles for quantitative analyses.

## II. Supplementary Discussions

### 1. Origins of different signals in visible TA spectra

In ZnO with low defect density, such as single-crystal films[2] and nanowires,[3] the TA responses in visible range are dominated by the photo-bleaching due to the hole saturation in defect-states. In contrast, the high defect density in single crystals,[4] nanobulks,[5] and nanoparticles,[6] exhibits pronounced photoinduced absorption features, which are originated from the carrier localization and self-trapping within a continuum of defect states. The carrier trapping opens up additional ultrafast relaxation channels, for example, continuum defect band diffusion,[4] trap-assisted hopping recombination,[5] and the formation of singlet/triplet mixed self-trapped excitons.[6]

## 2. Origins of the slow decays of band-edge PB

The slow relaxation of free exciton with characteristic time of $\tau_2^{3.4\ eV}$, is assigned to the radiative recombination, consistent with the 20–110 ps of radiative recombination observed in other ZnO nanomaterials.[7,8] In contrast to the first decay, the amplitude of this component decreases with increasing fluence, as tabulated in Table S4, indicating the competition of carrier relaxation channels through non-radiative Auger and radiative recombination.[9] The measured $PL_{UV}$ accounting for only 5.1% of the total emission intensity (see Figure S1d) also demonstrates that the radiative recombination is not dominant in our samples under high excitation fluence. Unlike the direct band-gap excitation, the extra carriers that are excited into the conduction band during three-body processes, such as hole Auger capturing[3] and defect-assisted Auger recombination,[10] eventually relax into the deep defect states, at where they recombine with the previously captured holes.[11] This relaxation gives rise to the measured $\tau_3^{3.4\ eV}$, which also serves as the transition time of band-edge PB to PIA2.

## 3. PIA2 originated from the defect-bound excitons

Photoinduced absorption signals usually correspond to interband transitions,[12,13] but PIA2 stems likely from the excited-state absorption of defect-bound excitons.[14,15] The large exciton binding energy of ZnO-NPs stabilizes the localized excitons against the thermal dissociation,[16] leading to a prerequisite of energy for probe photons to dissociate and excite these excitons to the band-edge states. Consequently, only the photons with energy exceeding the sum of the transition from defect levels and the exciton binding energy can stimulate the PIA2 signal.[17,18] The decay of PIA2 hence reflects the recombination of defect-bound excitons, which possess a long-lived lifetime more than 7 ns.[3,19] Similar lifetimes have been observed in the excitons trapped in deep-level defects.[2,20] In our case, the spin-orbit coupling induced by the lattice distortion around oxygen vacancies facilitates the otherwise forbidden triplet–singlet transitions, resulting in the nanosecond luminescence process.[6,20,21]

**4. Deviation from Debye-Waller effect indicating the local distortions**

When localized charge–lattice interactions induce asymmetric lattice distortions, those high-order diffraction peaks are sensitive to such distortions due to their small interplanar spacing.[22] The observed deviation is valid across various excitation fluences (see Figure S4), indicating that the residual local distortions may persist at thermal equilibrium. When only the first five peaks with low-order are considered, the derived root-mean-square atomic displacements follow the linear fluence-dependence of Debye-Waller effect, as depicted in the inset of Fig. 3c in the main text, testifying the global lattice thermalization through coupling to hot carriers.

**5. Correlation of local distortion with the Auger-induced thermalization**

The rise and decay times of $R(d/\text{int})$ gives a sum time of 10.3 ps, representing the total time of the local distortion motions preceding the global lattice thermalization. This characteristic time is close to the rapid thermalization given by the averaged intensity drops of (013) and (112) peaks (see the main text), suggesting that the local distortion is a concomitant of the rapid thermalization. The fluence-dependent measurements reveal that the rapid thermalization process is expedited with increasing fluence while the fit weight is increased, see Figure S6 and Table S7 for details. Such trends are consistent with the Auger heating mechanism,[23] in which the increased density of hot carriers in confined nanoparticles enhance the three-particle interactions, leading to exciton annihilations and the accelerated lattice heating.[24]

## 6. Anisotropic lattice responses as the precursor to the formation of small polarons

The distinct lattice motions of local distortion over the global thermalization also exhibit in the interplanar spacing changes, which are calculated using the centroid method, as depicted in Figure S7. During the early 10 ps, the (010) and (110) planes show anomalous contractions while the (002) and (013) planes show rapid expansions. Moreover, the expansion rates of (102) and (112) planes are apparently slower than those of (002) and (013) planes. Such transient structural responses are often associated with the formation of localized excited states, which induce the lattice deformations[25] that can be viewed as an initial step in stabilizing the excited-state small polarons.

## 7. Formation time of the oxygen-vacancy polarons

Time-resolved X-Ray absorption spectrum (TRXAS) measurement and *ab-Initio* calculations demonstrated that the formation of oxygen-vacancy polarons induced a local lattice distortion characterized by the damped oscillation with period of 300 fs,[11] which was not resolved in our UED results due probably to the limited temporal resolution. Instead, our UED results resolved the subsequent picosecond-scale expansion and relaxation dynamics, which manifested the stabilization of oxygen-vacancy polarons through local lattice distortions.

III. **Supplementary Figures**

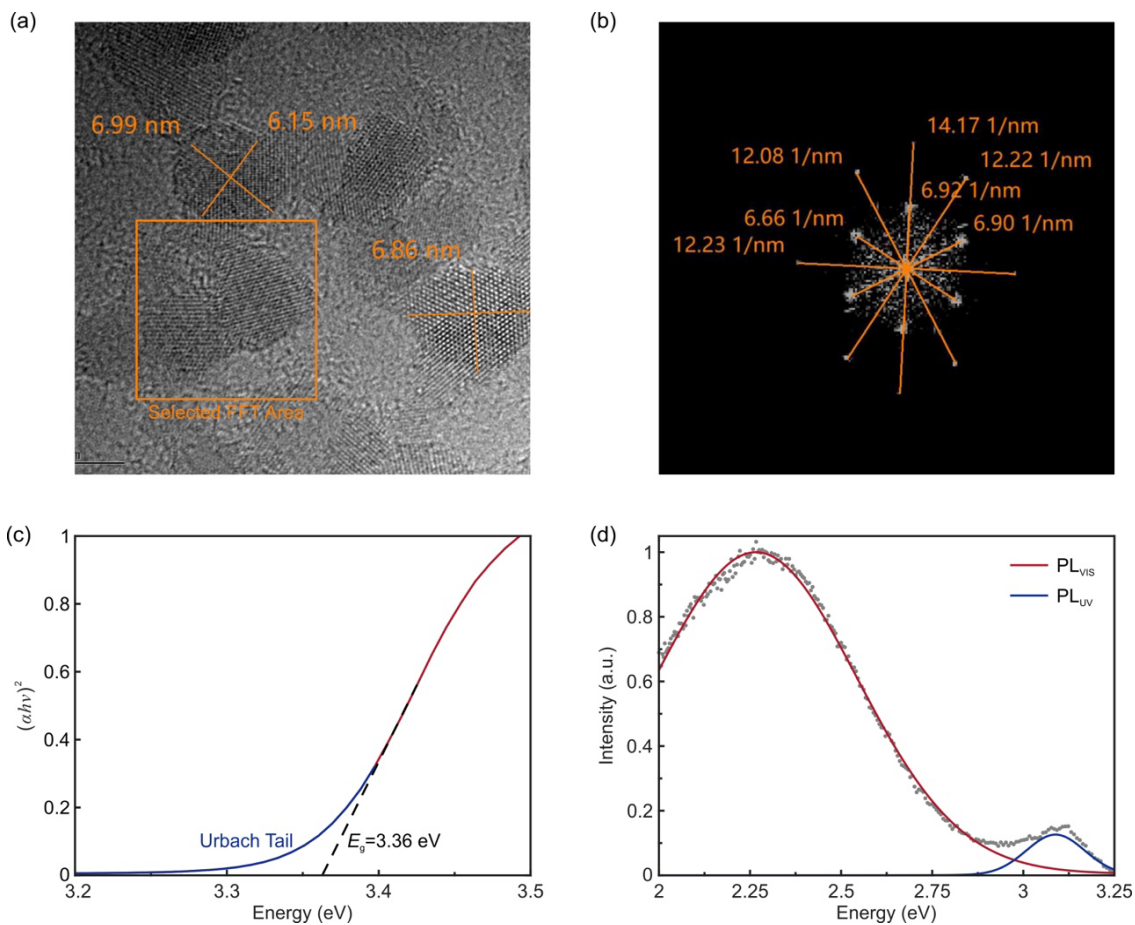

**Figure S1.** Sample characterizations. (a) HRTEM image. (b) Fast-Fourier transformation analysis of the selected area in (a). (c) Steady-state absorption spectrum. (d) Photoluminescence spectrum.

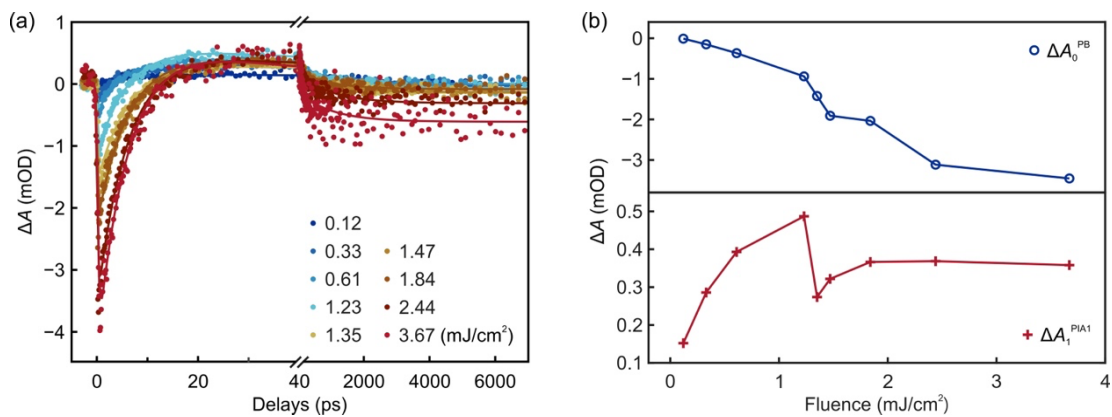

**Figure S2.** Fluence-dependent kinetics of the TA signal at 3.1eV. (a) Kinetic traces upon excitation of various fluences. Solid lines, multiexponential fits (See Table S5 for the fit results). (b) Upper panel: Dependence of the maximum amplitude of PB band at 3.1eV on excitation fluence, showing a saturation

at fluence of 2.44 mJ/cm² due to the governing Auger processes. Lower panel: Dependence of the maximum amplitude of PIA1 at delay time of 40 ps on fluence, extracted at 3.1eV whereby the contributions of both PB band and PIA1 overlap. The saturation of this mixed response occurs at 1.23 mJ/cm², the same fluence for the observed amplitude saturation of PIA2 at 3.4 eV (see the inset in Fig. 2b in the main text), suggesting a shared origin related to the defect-assisted carrier trapping.

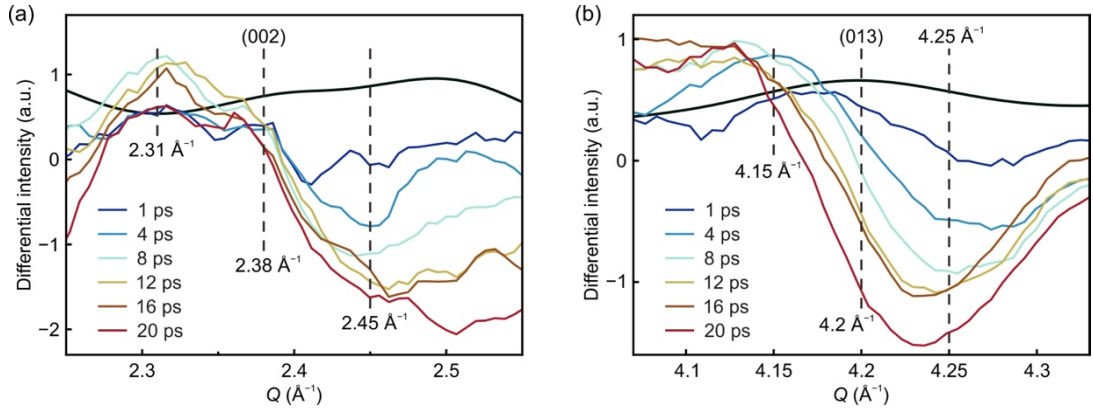

**Figure S3.** Temporal evolutions of differential diffraction intensity around (002) and (013) peaks. (a) Intensity change at 2.31 Å$^{-1}$ reaches its maximum at ~8 ps, indicating a significant left shift of the (002) peak, which intensity remains almost unchanged. The intensity change of the opposite-side shoulder of (002) peak at 2.45 Å$^{-1}$ exhibits slow dynamics due to the asymmetric peak profile and the overlapping with adjacent peaks. Such a derivative-like profile change of peak shift suggests the transient modification of local potential landscape, which is probably associated with the presence of short-lived lattice polarizations caused by hot carrier localization.[26] (b) Intensity change at 4.15 Å$^{-1}$ reaches its maximum at ~4 ps, accompanied with the slow dynamics of the opposite-side shoulder of (013) peak. Other crystallographic planes with normal vector exhibiting large projection along the *c*-axis show similar peak profile changes.

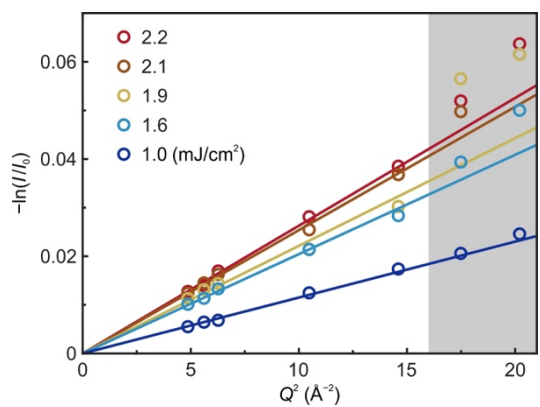

**Figure S4.** Dependence of the relative intensity changes at delay time of 1000 ps on squared scattering vector, upon excitation of various fluences. The high-$Q$ diffraction peaks, i.e., the (013) and (112) peaks, apparently deviate from the linear relationship described by the Debye-Waller effect, indicating the contributions of non-thermal structural change to the measured diffraction intensity.

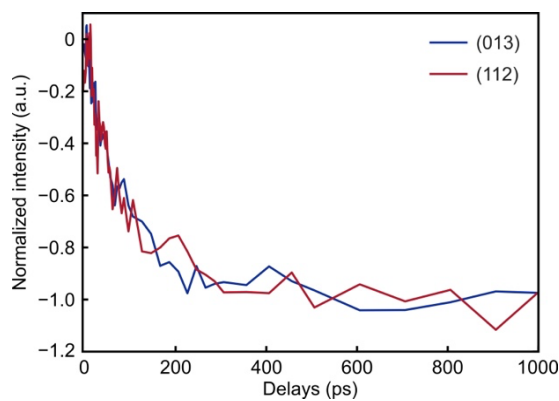

**Figure S5.** Normalized intensity changes of (013) and (112) peaks, showing identical temporal evolutions.

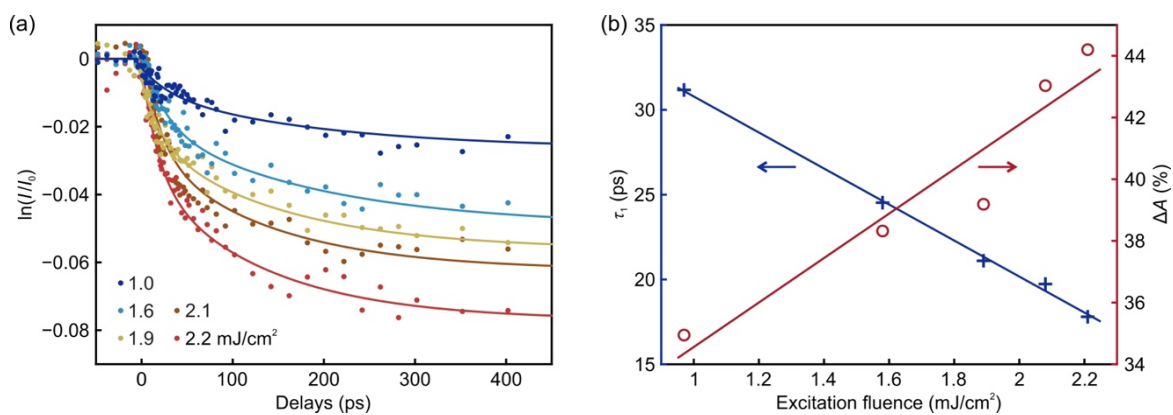

**Figure S6.** Intensity drops of (112) peak upon excitation of various fluences. (a) Kinetic traces. Solid lines, biexponential fits (See Table S7 for fit results). The biexponential fit yields two processes, a fast

component of a couple tens of picoseconds, which is associated with the Auger thermalization, and a slow component in the range of 130–190 ps, which is associated with the heat transfer towards substrate. (b) Dependences of the characteristic time and the fit weight of the fast component on excitation fluence, showing a trend consistent with Auger processes.

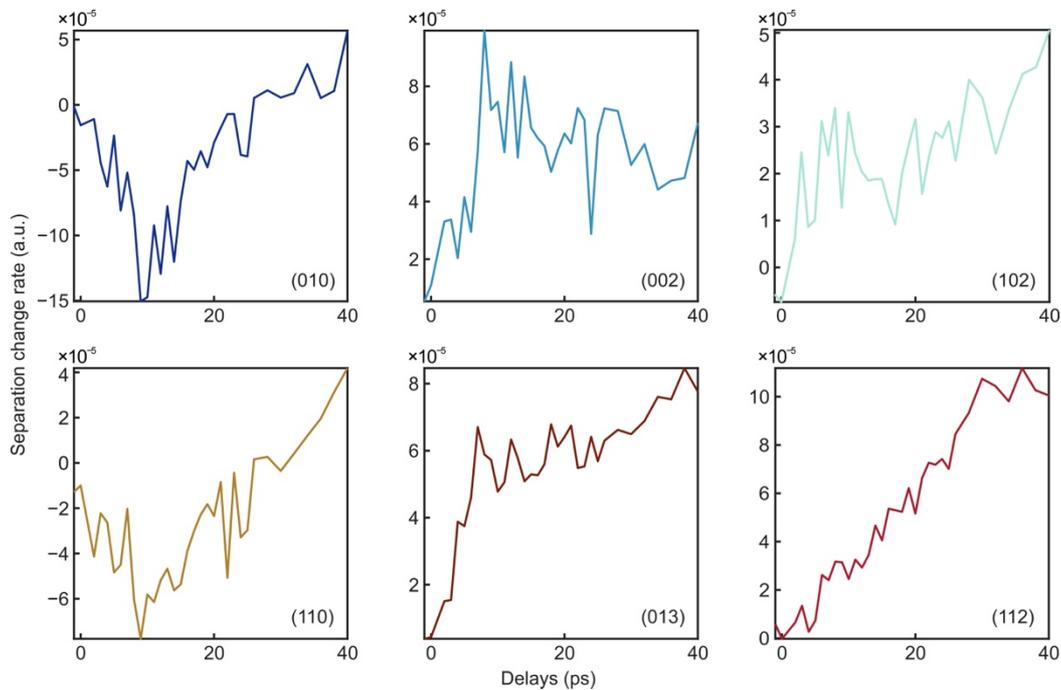

**Figure S7.** Change rates of lattice plane separation calculated by using the centroid method. In the early 10 ps after excitation, rapid expansions are observed in the (002) and (013) planes, coinciding with the extracted differential intensity changes shown in Figure S3. In contrast, the (010) and (110) planes show remarkable contractions in the same period, followed by thermal expansions. Other planes show gradual interplanar expansions over time.

## IV. Supplementary Tables

**Table S1.** Fit results of the PIA1 kinetic traces in Fig. 1c in the main text.

| Probe (eV) | 2.1 | 2.2 | 2.3 | 2.4 | 2.5 | 2.6 | 2.7 | 2.8 |
|---|---|---|---|---|---|---|---|---|
| $\tau_0^{PIA1}$ (ps) | 0.16 | 0.18 | 0.19 | 0.22 | 0.24 | 0.26 | 0.27 | 0.32 |
| $\Delta A_1^{PIA1}$ (%) | 48.7 | 43.4 | 43.8 | 43.6 | 40.5 | 40.1 | 38.4 | 32.5 |
| $\tau_1^{PIA1}$ (ps) | 5.71 | 5.17 | 5.99 | 6.57 | 6.44 | 7.43 | 8.45 | 8.55 |

**Table S2.** Fit results of the fluence-dependent kinetic traces of PIA1 in Fig. 1d in the main text.

| Fluence (mJ/cm²) | 0.12 | 0.33 | 0.61 | 1.23 | 1.35 | 1.47 | 1.84 | 2.44 | 3.67 |
|---|---|---|---|---|---|---|---|---|---|
| $\Delta A_0^{PIA1}$ (mOD) | 0.46 | 1.10 | 1.75 | 3.35 | 4.15 | 4.98 | 4.94 | 7.88 | 11.15 |
| $\tau_0^{PIA1}$ (ps) | 0.16 | 0.15 | 0.14 | 0.16 | 0.17 | 0.15 | 0.16 | 0.15 | 0.14 |

**Table S3.** Fit results of the energy-dependent kinetic traces of PB band in Fig. 2a in the main text.

| Probe energy (eV) | 3.1 | 3.15 | 3.2 | 3.25 | 3.3 | 3.4 |
|---|---|---|---|---|---|---|
| $\Delta A_0^{PB}$ (mOD) | −0.64 | −1.14 | −2.05 | −2.85 | −3.81 | −5.09 |
| $\tau_0^{PB}$ (ps) | 0.15 | 0.14 | 0.11 | 0.19 | 0.12 | 0.17 |
| $\Delta A_1^{PB}$ (mOD) | 1.19 | 1.67 | 2.46 | 3.21 | 4.08 | 4.28 |
| $\tau_1^{PB}$ (ps) | 3.8 | 5.6 | 6.2 | 7.5 | 9.1 | 10.9 |
| $\Delta A_2^{PB}$ (mOD) | −0.59 | −0.55 | −0.45 | −0.36 | −0.14 | 2.12 |
| $\tau_2^{PB}$ (ps) | 278.5 | 279.4 | 280 | 281.2 | 280 | 279.9 |

**Table S4.** Fit results of the fluence-dependent kinetic traces of band-edge PB in Fig. 2b in the main text.

| Fluence (mJ/cm²) | 0.12 | 0.33 | 0.61 | 1.23 | 1.35 | 1.47 | 1.84 | 2.44 | 3.67 |
|---|---|---|---|---|---|---|---|---|---|
| $\Delta A_0^{3.4\,eV}$ (mOD) | −1.27 | −2.65 | −3.78 | −6.20 | −7.19 | −8.27 | −8.44 | −9.94 | −9.30 |
| $\tau_0^{3.4eV}$ (ps) | 0.23 | 0.17 | 0.14 | 0.16 | 0.15 | 0.15 | 0.17 | 0.16 | 0.11 |
| $\Delta A_1^{3.4eV}$ (%) | 26.1 | 33.8 | 41.6 | 49 | 52.3 | 53.1 | 54 | 59.7 | 75.8 |
| $\tau_1^{3.4eV}$ (ps) | 13.6 | 12.2 | 10.8 | 9.4 | 8.4 | 7.3 | 6.8 | 6.9 | 6.3 |
| $\Delta A_2^{3.4eV}$ (%) | 34.2 | 25.4 | 25.5 | 21.1 | 25.5 | 26.3 | 25.9 | 25.6 | 12.5 |
| $\tau_2^{3.4eV}$ (ps) | 58.5 | 57 | 55.5 | 54 | 52.6 | 49.3 | 46.9 | 36.9 | 46.6 |
| $\Delta A_3^{3.4eV}$ (%) | 49.7 | 40.8 | 32.9 | 29.9 | 22.2 | 20.6 | 20 | 14.6 | 11.7 |
| $\tau_3^{3.4eV}$ (ps) | 492 | 484 | 475 | 468 | 454 | 445 | 432 | 421 | 434 |
| $\Delta A_1^{PIA2}$ (mOD) | 0 | 0 | 0 | −0.59 | −0.52 | −0.51 | −0.6 | −1.42 | −1.04 |
| $\tau_1^{PIA2}$ (ps) | inf | inf | inf | 7005 | 8621 | 8477 | 8974 | 9847 | 7206 |

**Table S5.** Fit results of the fluence-dependent kinetic traces of PB band at 3.1 eV in Figure S2.

| Fluence (mJ/cm²) | 0.12 | 0.33 | 0.61 | 1.23 | 1.35 | 1.47 | 1.84 | 2.44 | 3.67 |
|---|---|---|---|---|---|---|---|---|---|
| $\Delta A_0^{3.1\,eV}$ (mOD) | −0.087 | −0.334 | −0.599 | −1.282 | −1.853 | −2.411 | −2.582 | −3.885 | −4.509 |
| $\tau_0^{3.1eV}$ (ps) | 0.1 | 0.19 | 0.1 | 0.1 | 0.1 | 0.1 | 0.11 | 0.1 | 0.15 |
| $\Delta A_1^{3.1eV}$ (mOD) | 0.23 | 0.621 | 0.975 | 1.799 | 2.237 | 2.788 | 2.935 | 4.266 | 4.913 |
| $\tau_1^{3.1eV}$ (ps) | 3 | 4.03 | 4.99 | 5.69 | 6.05 | 6.31 | 6.65 | 6.73 | 6.92 |
| $\Delta A_2^{3.1eV}$ (mOD) | −0.14 | −0.26 | −0.36 | −0.532 | −0.45 | −0.458 | −0.433 | −0.71 | −0.969 |
| $\tau_2^{3.1eV}$ (ps) | 296.2 | 288.7 | 284.3 | 280.1 | 275.9 | 269.5 | 263.8 | 261 | 254.3 |

**Table S6.** Fit results of the intensity drop kinetics of the five selected diffraction peaks in Fig. 3b in the main text.

| Planes | (010) | (011) | (110) | (013) | (112) |
|---|---|---|---|---|---|
| $A_1$ (a. u.) | 0.0033 | 0.0058 | 0.0131 | 0.0192 | 0.0217 |
| $\tau_1$ (ps) | 13.9 | 15.6 | 16.9 | 22.1 | 17.8 |
| $A_2$ (a. u.) | 0.0089 | 0.0108 | 0.0242 | 0.0305 | 0.0399 |
| $\tau_2$ (ps) | 95.0 | 85.2 | 82.3 | 119.7 | 129.4 |

**Table S7.** Fit results of the intensity drop kinetics of (112) peak in Figure S6.

| Fluence (mJ/cm²) | 1.0 | 1.6 | 1.9 | 2.1 | 2.2 |
|---|---|---|---|---|---|
| $A_1$ (%) | 34.9 | 38.2 | 39.2 | 43.0 | 44.2 |
| $\tau_1$ (ps) | 31.2 | 24.5 | 21.1 | 19.7 | 17.8 |
| $A_2$ (%) | 65.1 | 61.7 | 60.8 | 57.0 | 55.8 |
| $\tau_2$ (ps) | 193.5 | 183.4 | 146.7 | 134.5 | 129.4 |